\documentclass[showpacs,nofootinbib,aps,preprintnumbers,
floatfix,superscriptaddress,prl,twocolumn]{revtex4}
\usepackage{graphicx} 
\usepackage{epsfig}
\usepackage{psfrag}
\usepackage{dcolumn} 
\usepackage{bm} 
 
\def\gsim{\lower0.5ex\hbox{$\:\buildrel >\over\sim\:$}}
\def\lsim{\lower0.5ex\hbox{$\:\buildrel <\over\sim\:$}}
 
\newcommand{\bea}{\begin{eqnarray}}
\newcommand{\eea}{\end{eqnarray}}
 
\begin{document}
 
\preprint{CUMQ/HEP 136 }
\preprint{HIP-2005-32/TH}
\preprint{hep-ph/ddmmyyy}
 
\title{Spontaneous CP and R parity breaking in supersymmetry}
 
\author{M. Frank} 
\email{mfrank@vax2.concordia.ca} 
\affiliation{Department of Physics, Concordia University,
7141 Sherbrooke St. West, Montreal, Quebec, Canada H4B 1R6}
\author{K. Huitu}
\email{Katri.Huitu@helsinki.fi}
\affiliation{High Energy Physics Division, Department of Physical
Sciences, P.O. Box 64, FIN-00014 University of Helsinki, Finland}
\affiliation{Helsinki Institute of Physics,
P.O. Box 64, FIN-00014 University of Helsinki, Finland}
\author{T. R\"uppell}
\email{timo.ruppell@helsinki.fi}
\affiliation{High Energy Physics Division, Department of Physical
Sciences, P.O. Box 64, FIN-00014 University of Helsinki, Finland}
\affiliation{Helsinki Institute of Physics,
P.O. Box 64, FIN-00014 University of Helsinki, Finland}

\received{\today}
 
\pacs{12.60.Jv, 11.30.Er, 14.60.Pq, 11.30.Qc}

\begin{abstract} \vspace*{10pt}
We show that a model where both CP and R-parity are spontaneously
broken exists.  
We study the electroweak symmetry breaking sector of the model
and find minima consistent with experimentally viable Higgs boson 
masses.
We also demonstrate that one can obtain neutrino masses and mixing
angles within measured values.
\end{abstract}

\maketitle
 
Breaking CP 
in the Higgs sector, either spontaneously or explicitly,
is a theoretically attractive alternative to the CP violating phase 
present in the Standard Model (SM) \cite{Lee, Weinberg}. 
New sources for CP violation from beyond the
Standard Model scenarios
are needed to generate enough CP violation 
to explain the matter-antimatter asymmetry in the universe.
While in supersymmetric models a large number of new phases
emerge, in a general minimal supersymmetric standard model (MSSM) these phases 
are strongly constrained by electric dipole moments \cite{AKL}.

A defect of the Standard Model, which persists in the MSSM, is that the neutrino masses vanish.
  Yet the neutrino experiments have provided strong
evidence for small nonvanishing neutrino mass \cite{PDG}.
One popular way to explain the neutrino masses is a small violation of
R-parity \cite{HS}, $R_p=(-1)^{3B-L+2s}$, where $B$=baryon number,
$L$=lepton number, and $s$=spin of the particle. Another is the seesaw mechanism
\cite{seesaw} which  can generate small masses for neutrinos 
by allowing Majorana masses through the presence of heavy right-handed neutrinos.

Supersymmetric models share the problem of origin of CP violation with
the Standard Model.
In addition, there is no fundamental reason for the existence of R-parity in the MSSM, where it
 is put in by hand in order to protect the proton from
decaying.
However, if only lepton number (or baryon number) is violated, the proton
does not decay.

Both neutrino masses and CP violation could be explained if CP and R were symmetries
of the Lagrangian, but spontaneously broken by the vacuum. 
In this letter we provide a model where both of these violations are
intertwined and both are spontaneous.
We show that while it is non-trivial to satisfy conditions for both symmetry breakdowns at the
 same time, there are regions in the parameter space where we can realize
suitable Higgs masses, as well as measured neutrino mass differences 
and mixing angles.

A model where both favored neutrino mass generation mechanisms
- namely seesaw and R-parity violation -
with spontaneous R-parity violation was realized in \cite{KO}, where  
 the spontaneous $R_p$ violation 
was introduced via a term proportional to $ N L H_2$.
This represents the familiar bilinear $R_p$ violating term
mixing lepton and Higgs  superfields, 
$L H_2$, when the right-handed sneutrino field, $\tilde N$, 
develops a VEV. 
This term also  breaks the lepton number  spontaneously (but not R-parity), and 
thus introduces a superfluous massless Goldstone
boson into the scalar spectrum. 
The problem can be solved by adding a singlet $S$ to the
theory, through a term $N^2 S$,
which explicitly breaks lepton number, as $S$ cannot be assigned
lepton number~$-2$.

Attempts to violate CP spontaneously, by complex VEVs of the 
neutral scalars exist  \cite{Lee},
but fulfilling the experimental
constraints has proven difficult.
More than one Higgs doublet is needed, see {\it e.g.} 
\cite{BLS}. Spontaneous breaking of CP is not possible at tree level in
the MSSM with two Higgs doublets, while it is allowed in
a model with three doublets  
\cite{Branco}.
Instead of adding doublets, one can study extended models, like
the NMSSM model  \cite{NSW}, where the so called $\mu$-problem 
has been avoided by
adding a singlet and requiring $\mathbf Z_3$ symmetry. 
At tree-level one cannot get spontaneous CP
violation in this model either and consequently
radiative corrections were evoked \cite{HJO}.
In that case a very light Higgs boson emerges \cite{GP},
as also happens in the MSSM, if spontaneous CP violation is induced
via radiative correction \cite{MP}.
Another possibility studied is to discard the $\mathbf Z_3$ symmetry
completely.  
On one hand, this way one loses the solution to the $\mu$ problem, 
on the other hand, it is possible to achieve SCPV \cite{BKRT} and
also solve the problem of domain walls, which
are created during the EW phase transition as the $\mathbf{Z}_3$
symmetry is broken spontaneously.

An interesting model for spontaneous CP violation was presented
in \cite{HRT}, where the  $\mathbf{Z}_3$ symmetry is replaced 
by R-symmetries on the whole superpotential, including non-renormalizable terms
\cite{PT}.
The method generates a tadpole term for the singlet
field $S$ in the soft SUSY breaking part of the Lagrangian. This tadpole
term
allows for spontaneous CP violation to occur at tree-level 
\cite{HRT,bib:hep-ph/9803271}.
The tadpole is assumed to originate from non-renormalizable
interactions, which do not spoil quantum stability.
We adopt this approach here.

The superpotential of our model is
\bea
W &=& h_U^{ij}Q_i H_2 U_j
   -h_D^{ij}H_1 Q_i D_j
   -h_E^{ij}H_1 L_i E_j \nonumber\\
&+& h_N^{ij}L_i H_2 N_j 
   +\lambda_H H_1  H_2 S
   +{\lambda_S \over 3!} S^3
   +{\lambda_{N_i} \over 2} N_i^2 S, \nonumber
\eea
where $H_1$ and $H_2$ denote the Higgs doublet superfields, $L_i$ and $Q_i$
the left-handed lepton and quark doublet superfields, 
respectively, and $E_i$ and $U_i$, $D_i$ the lepton and quark singlet
superfields.
Right-handed neutrino superfields are denoted by $N_i$ and $S$ is the
gauge singlet superfield.
The terms in the Lagrangian are the only renormalizable ones
that respect CP and $R$-parity, in addition to the gauge
symmetry. As all the parameters in the Lagrangian are real, this solves the
strong CP problem \cite{BLS}. 
The only possible global symmetries are the baryon number and
$\mathbf Z_3$.
 
The soft SUSY breaking terms in this model are the mass terms for
scalars and gauginos, and the part
mirroring the superpotential with an additional $S$-tadpole,
 \bea
 V_{soft} &=&
     \Big[ A_U^{ij}\tilde Q_i H_2 \tilde U_j
    -A_D^{ij} H_1 \tilde Q_i \tilde D_j
    -A_E^{ij} H_1 \tilde L_i \tilde E_j \nonumber\\
&+&  A_N^{ij}\tilde L_i H_2 \tilde N_j
    +A_H H_1 H_2 S
    +{A_S \over 3!} S^3
    +{A_{N_i} \over 2} \tilde N_i^2 S \nonumber\\
&-& \xi^3 S + h.c.\Big] + M_{\phi}^{2\ ij} \phi_i^\ast\phi_j 
    + M_a (\lambda_a^2 + \lambda_a^{\ast 2}). \nonumber
\eea
 Here $\phi$ runs over the scalar fields, $i,j$ over the possible family
indices and $a$ over the three gauge groups. In calculations, we
take the mass matrices to be flavor diagonal, and
assume that only the third generation
Yukawa couplings (except those for neutrinos) differ from zero.
We then impose the same
texture on the corresponding $A$-terms, and treat the neutrino Yukawa couplings as
free parameters.
The full tree-level scalar potential is $V_s=V_{soft}+V_F+V_D$,
where $V_F$ and $V_D$ are the usual $F$ and $D$ terms.

The minimization of the scalar potential with respect to the 
fields $\phi_i$ and the corresponding phases $\theta_i$
yield constraints later used in finding the scalar mass matrix,
\begin{eqnarray}
\label{eqn:vac}
\left. {\partial V_s \over
\partial\phi_i}\right|_{\phi=\langle\phi\rangle}
 = 0,\;\;
 \left. {\partial V_s \over
\partial\theta_i}\right|_{\phi=\langle\phi\rangle}
= 0.
\end{eqnarray}
Without spontaneous CP violation, the VEVs are real and the
minimization equations with respect to the phases are always satisfied. 

The minimization equations for the charged scalars can be trivially
solved by setting all charged scalar VEVs to zero. 
As long as the tree-level masses of these fields
remain positive and the corresponding soft $A$-terms remain small
enough, this is also the global minimum of the potential 
with respect to these fields \cite{KO}.
We use the seventeen equations (the phase of 
the $H_1$ field can always be rotated away) given by the moduli and phases of 
the neutral scalar fields to solve for the
soft masses of the corresponding fields and subset of $A$-parameters 
($A_N,\,A_S,\,A_{N_i},\,A_N^{i,3}$). Thus the complex VEVs remain free parameters and we denote
\begin{eqnarray}
\nonumber
&&\langle H_1 \rangle = \left(\begin{array}{c}v_1\\0\end{array}\right),\;
\langle H_2 \rangle =  \left(\begin{array}{c}0\\v_2
    e^{i\delta_2}\end{array}\right), \;
\langle S \rangle = \sigma_S e^{i\theta_S},\\
&&\langle \tilde L_i \rangle =  \left(\begin{array}{c}\sigma_{L_i}
    e^{i\theta_{L_i}}\\0\end{array}\right),\;
\langle \tilde N_i \rangle = \sigma_{R_i} e^{i\theta_{R_i}}.
\end{eqnarray}
Note that since $R_p$ is violated, the $W$ mass is $m_W^2= {1\over 2}
g_2^2 v^2$
where $ v^2 \equiv v_1^2
+ v_2^2 + \sigma_{L_i}^2 \approx (174$ GeV$)^2$. 

Writing the fields as
$\phi \equiv \phi_r + i \phi_i$, with nine neutral scalar fields (two Higgs, one singlet and six
sneutrinos) we get  
an 18$\times$18 dimensional mass matrix for the scalars.
The radiative corrections to the scalar masses are implemented
via the one-loop effective scalar potential,
\begin{eqnarray}
V_{1-loop}&=&{-3\over 32 \pi^2}\left[
m^4_{\tilde t_1}\log\left({m^2_{\tilde t_1}\over \Lambda^2}-{3\over 2}\right)\right.\nonumber\\ 
&+&\left.m^4_{\tilde t_2}\log\left({m^2_{\tilde t_2}\over \Lambda^2}-{3\over 2}\right)
-2m^4_t\log\left({m^2_t\over\Lambda^2}-{3\over 2}\right)\right],\nonumber
\end{eqnarray}
where $\Lambda$ is the renormalisation scale. 
A similar term for the bottom (s)quark masses 
contributes significantly, if $\tan\beta$ is large. Here
$m_t^2=
y_t v_2^2$ and $m^2_{\tilde t_i}$ are  
the eigenvalues of the stop mass matrix,
\bea
M^2_{\tilde t_L\tilde t_L}&=&M^2_{Q_{33}}+m_t^2
      -({g_1^2\over 12}-{g_2^2\over 4})(v_1^2-v_2^2+\sigma_{L_i}^2),\nonumber\\
M^2_{\tilde t_L\tilde t_R}&=& A_t v_2 e^{-i\delta_2} \nonumber\\
&+& y_t (v_1 \lambda_H \sigma_S e^{i\theta_S} 
+h_N^{ij}\sigma_{L_i}\sigma_{R_j}e^{i(\theta_{L_i}-\theta_{R_j})}),\nonumber\\
M^2_{\tilde t_R\tilde t_R}&=&M^2_{U_{33}}
               +m_t^2+{g_1^2\over 3}(v_1^2-v_2^2+\sigma_{L_i}^2).
\eea
The loop corrections lead to additional terms in both the
minimisation conditions and the scalar mass matrix.  In numerical
calculations, we omit the D-term contributions and set for simplification, 
$M_{Q_{33}}=M_{U_{33}}=M_{SUSY}$, with $M_{SUSY}\sim \Lambda\sim 1$ TeV.

The following experimental input is used
\begin{eqnarray} \nonumber
&& v^2 = 174\; \mathrm{ Gev},\;
m_W = 80.42\; \mathrm{ GeV},\;
m_t^{Pole} = 180\; \mathrm{ GeV},\;\nonumber\\
&&\alpha_s = 0.102,\;
m_\tau = 1.777\; \mathrm{ GeV},\; \sin^2\theta_w = 0.23124
\end{eqnarray}
The rest of our free parameters are randomly sampled, with 
sampling ranges as follows (the couplings $\lambda_i$ are constrained by
perturbativity):
\bea
&&0.1<\lambda_{H,N_i}<0.4,\;\;0.2<\lambda_S<0.7,\;\;
|h_N|<10^{-7},\nonumber\\
&&0.4 \;{\rm TeV}< \xi <1\;{\rm TeV},\;\;
-\pi<\theta_{\phi}<\pi,\;\;
|\langle S\rangle|<1\;{\rm TeV},\;\;\nonumber\\
&&|\langle \tilde\nu_L\rangle|<100\;{\rm keV},\;\;
|\langle \tilde N_i\rangle|<1\;{\rm TeV},\;\;
2<\tan\beta<60,
\eea
and the $A$-parameters not eliminated by Eq.(\ref{eqn:vac}) vary between
$0<A_N^{ij}<(1\;{\rm TeV}) h_N^{ij}$.
 
In the limit $\lambda_{N_i} \to 0$ we recover lepton number conservation.
CP and R-parity are still spontaneously violated,
and thus in this limit the model contains an experimentally
disallowed Goldstone boson.
\begin{figure}[tpb]
\begin{center}
\mbox{\epsfig{file=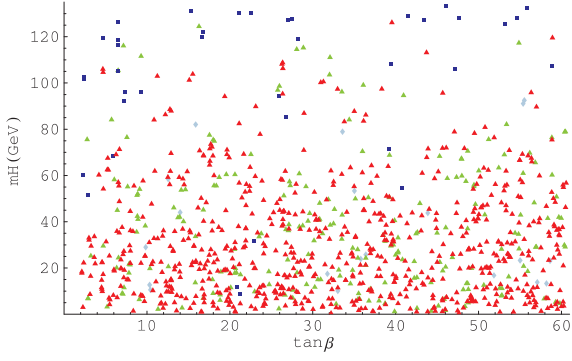, width=7.5truecm
}}
\end{center}
\caption{
Lightest physical scalar mass as a function of $\tan\beta$.
The main component of the lightest scalar is $H_{1,2}^0$ (blue
rectangle), $S$ (light blue diamond), $\tilde\nu_L$ (green triangle),
or $\tilde N$ (red triangle).}
\label{Higgs}
\end{figure}

There are a total of 55 relevant parameters $\langle \sigma_\phi
\rangle,\ \theta_\phi,\ \lambda_S,\ \lambda_H,\ \lambda_{N_i},
\ h_N^{ij},\ M_\phi^2,\ A_S,\ A_H,\ A_{N_i},$ $ A_N^{ij}$ and $\xi$. $17$
of these can be eliminated by using the minimization
conditions, leaving 38 free parameters. Sampling over such a large space
makes it extremely difficult to find viable
minima of the scalar potential, even without applying any other cuts or
considerations.
For the full model we choose to set
$\sigma_{R_{1,2}}=\theta_{R_{1,2}}=0$ to further reduce the sampling
space. This choice identically solves
the vacuum conditions $\partial_{\theta_{R_1}}V=0$ and
$\partial_{\theta_{R_2}}V=0$.

In this model, the couplings of the lightest scalar are reduced compared
to the SM, due to the (sometimes dominant) singlet components.
Thus the corresponding experimental limits are smaller as well than the
experimental lower bound for the SM Higgs boson, $m_H\gsim 114$ 
GeV, see e.g. discussion in \cite{redbounds}.
In Fig. \ref{Higgs} we plot the mass of the
lightest Higgs boson as a function of $\tan\beta$, and show that
experimentally acceptable Higgs bosons masses can be found 
for a large part of the parameter space.

In a field basis of $\nu_{L_i},N_i,\tilde S,
\tilde H_1^0, \tilde H_2^0, \lambda_0, \lambda_3$ the neutral fermions
form the following 11$\times$11 mass matrix:
\begin{widetext}
\begin{equation}
M_{\chi^0} = \left(\begin{array}{ccccccc}
\mathbf{0}_{3\times 3} &
\mathbf{h}_N^{3\times3} \langle H_2^0\rangle &
\mathbf{0}_{3\times 1} &
\mathbf{0}_{3\times 1} &
h_N^{i,j} \langle\tilde N_{j}^\ast\rangle &
- {g_1\over \surd 2} \langle\tilde\nu_{L_i}^\ast\rangle &
{g_2\over \surd 2} \langle\tilde\nu_{L_i}^\ast \rangle \\
\mathbf{h}_N^{3\times 3} \langle H_2^0\rangle &
\mathbf{1}_{3\times 3}\lambda_{N_i} \langle S\rangle &
\lambda_{N_i} \langle\tilde N_{i}^\ast\rangle &
0 &
h_N^{j,i}\langle\tilde\nu_{L_j}\rangle &
0 &
0 \\
\mathbf{0}_{1\times3} &
\lambda_{N_i} \langle\tilde N_{i}^\ast\rangle &
\lambda_S \langle S\rangle &
\lambda_H \langle H_2^0\rangle &
\lambda_H \langle H_1^0\rangle &
0 &
0 \\
\mathbf{0}_{1\times3} &
0 &
\lambda_H  \langle H_2^0\rangle &
0 &
\lambda_H \langle S\rangle &
- {g_1\over \surd 2}  \langle H_1^0\rangle &
{g_2\over \surd 2} \langle H_1^0\rangle \\
h_N^{i,j} \langle\tilde N_{j}^\ast\rangle &
h_N^{j,i}\langle\tilde\nu_{L_j}\rangle  &
\lambda_H \langle H_1^0\rangle &
\lambda_H \langle S\rangle &
0 &
 {g_1\over \surd 2}  \langle H_2^{0\ast}\rangle &
- {g_2\over \surd 2} \langle H_2^{0\ast}\rangle \\
- {g_1\over \surd 2} \langle\tilde\nu_{L_i}^\ast\rangle &
0 &
0 &
- {g_1\over \surd 2} \langle H_1^0\rangle &
{g_1\over \surd 2} \langle H_2^{0\ast}\rangle &
M_1 & 0 \\
{g_2\over \surd 2} \langle\tilde\nu_{L_i}^\ast\rangle &
0 &
0 &
{g_2\over \surd 2} \langle H_1^0\rangle &
- {g_2\over \surd 2} \langle H_2^{0\ast}\rangle &
0 &
M_2
\end{array}\right)
\end{equation}
\end{widetext}
It is easy to see the structure of the usual seesaw mechanism, which
produces small neutrino masses $m_\nu$,
\begin{equation}
M_{\chi^0} = \left(\begin{array}{cc}
0   & m_D\\
m_D^T & M_R \end{array}\right), \quad m_\nu= 
-m_D M_R^{-1} m_D^T,
\end{equation}
where $ m_D \ll M_R$. Similarly to \cite{KO},
there are actually several sources for neutrino masses: the usual seesaw and
the mixing of neutrinos with $\tilde{h}_2^0$ and gauginos. 

Inspecting the requirement that $m_D\ll M_R$ yields some qualitative
understanding of the model. In particular,
the left-handed sneutrino VEVs must be
 small and $h_N \langle \tilde N^\ast\rangle$ should be of the same
order. Thus, although $\langle \tilde N^\ast\rangle$
is not bound by any other prior consideration, having $h_N\approx
10^{-7}$ results in an upper limit of a few TeV
for the right-handed sneutrino VEVs.
 
We diagonalise $M_{\chi^0}$ numerically and use $M_1\sim M_2\sim 1$ TeV. 
Great care must be taken, as the elements of
$M_{\chi^0}$ may vary over ten orders of magnitude, and the eigenvalues
themselves over as much as twenty orders of magnitude.
The diagonalising matrix $\mathcal{N}$, with
$\mathcal{N} M_{\chi^0} \mathcal{N}^{-1} =
\mathrm{diag}(m_{\chi^0_i},m_{\nu_j})$,
has the following general form
\begin{equation}
\mathcal{N}=\left(\begin{array}{cc}
\zeta&N_\chi\\V_\nu^T&\bar\zeta^T\end{array}\right).
\end{equation}
Here $\zeta,\bar\zeta \ll 1$ denote $8\times3$ matrices that can be
determined perturbatively, see e.g.
\cite{bib:PertDiag}.
Our interest lies in the matrix $V_\nu$, the neutrino mixing
matrix.
Using the canonical notation for the neutrino mixing matrix \cite{bib:PertDiag},
we can extract the mixing angles as follows
\begin{eqnarray}
\nonumber
&&\sin\theta_{13}=\left|V_\nu^{13}\right|
                = \left|\mathcal{N}^{11\ 1}\right|,\nonumber\\
&&\tan\theta_{12}=\left|{V_\nu^{12}\over V_\nu^{11}}\right|
                = \left|{\mathcal{N}^{10\ 1}\over \mathcal{N}^{9\
1}}\right|,\nonumber\\
&&\tan\theta_{23}=\left|{V_\nu^{23}\over V_\nu^{33}}\right|
                = \left|{\mathcal{N}^{11\ 2}\over \mathcal{N}^{11\
3}}\right|
\end{eqnarray}
 
\begin{figure}[tpb]
\begin{center}
\mbox{\epsfig{file=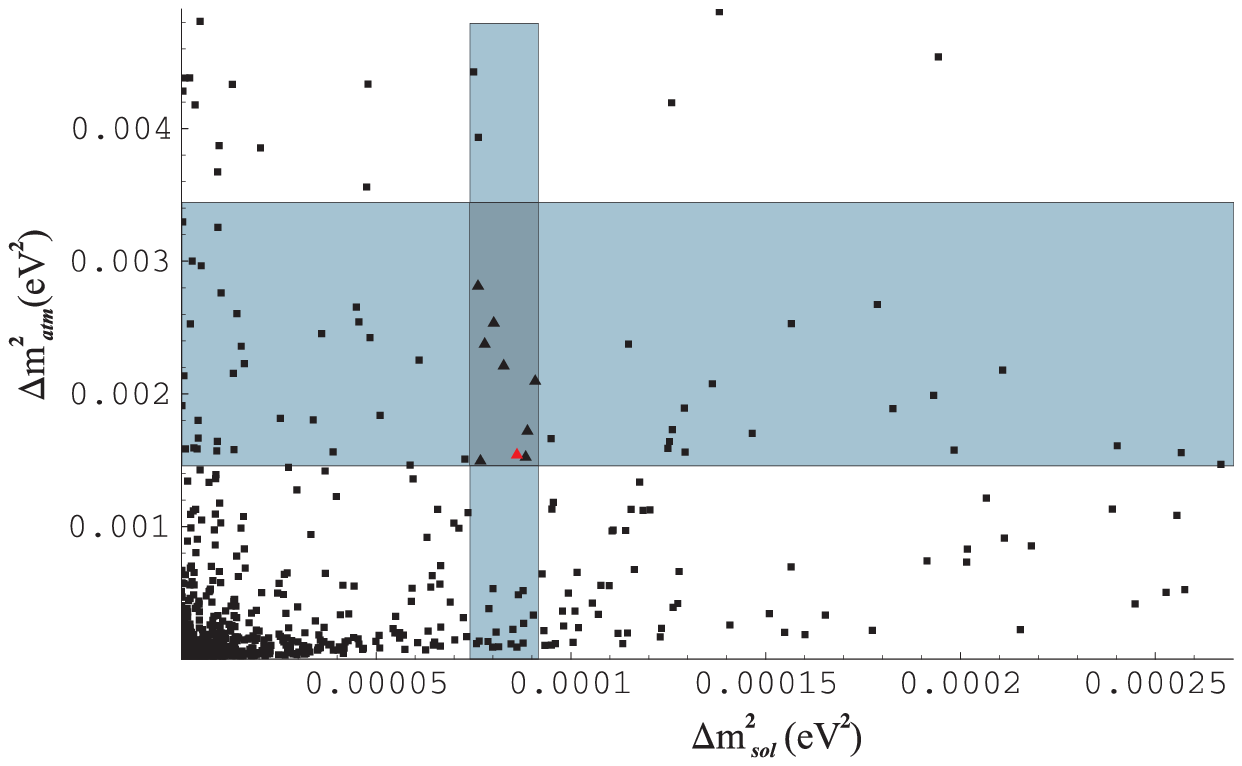, width=7.5truecm
}}
\mbox{\epsfig{file=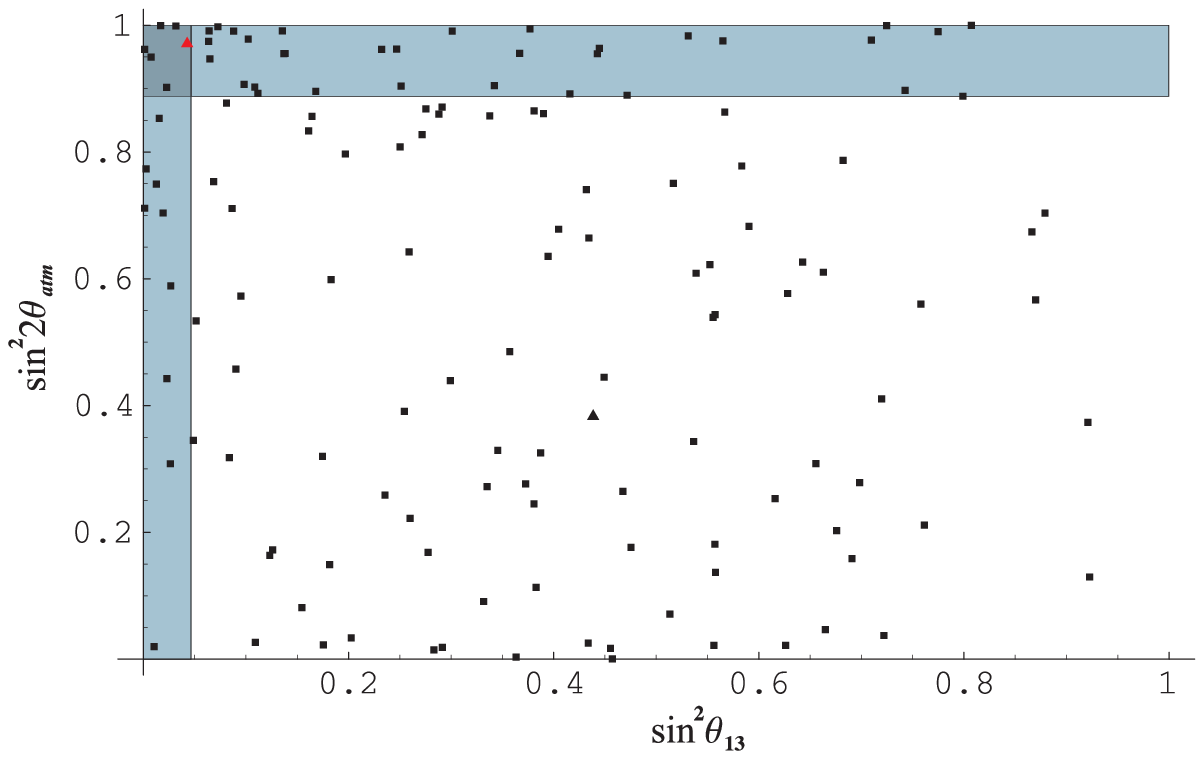, width=7.5truecm
}}
\end{center}
\caption{Correlations of neutrino angles and mass differences.
Experimentally allowed regions at 3$\sigma$ level are colored.
Points satisfying neutrino mass constraints are marked as triangles.
Upper plot: $\Delta m_{atm}^2$ vs  $\Delta m_\odot^2$,
lower plot: $\sin^2 2\theta_{23}$ vs $\sin^2 \theta_{13}$,
requiring at the same time that the experimental limit for 
$\sin^2 \theta_{12}$ is satisfied. One point satisfies all constraints (red).
}
\label{neutrinos}
\end{figure}

Once we have sufficient viable minima, i.e. points in parameter space
that produce a minimum of the potential, we  
apply experimental constraints.
We calculate how much each minimum  
deviates from the current
experimental results and select a number of closest events for further
analysis. The experimental input we use is  
the following \cite{bib:hep-ph/0411131}:
\bea
&&\sin^22\theta_{23}\ge 0.89,\;\; 
\sin^2\theta_{13}\le 0.047,\nonumber\\
&&\sin^2\theta_{12}\simeq 0.23 - 0.37,\nonumber\\
&&\Delta m_{atm}^2 \simeq 1.4\times 10^{-3} \mathrm{eV}^2 - 3.3\times
10^{-3} \mathrm{eV}^2,\nonumber\\
&&\Delta m_\odot^2 \simeq 7.3\times 10^{-5} \mathrm{eV}^2 -
9.1\times 10^{-5} \mathrm{eV}^2.
\eea
The first stage of finding points for the full model which are minima of the
scalar potential has a signal-to-noise ratio of one over fifty thousand, which shows the odds of having an event fall within the constraints. Of these points
only roughly one in a thousand will satisfy all the above constraints.
In Fig. \ref{neutrinos} we demonstrate that it is possible to
obtain models where the mass differences and experimentally 
found mixing angles are satisfied.

Summarizing, we have shown that a model that violates both CP invariance and
R-parity exists, and we constructed it explicitly. We have shown that
experimentally viable neutrino and Higgs boson masses can be obtained. 
We have reason to expect  that in our model EDM bounds and experimental 
results on kaon (especially
$\epsilon_K$) and B physics can be satisfied \cite{FHR}. 

\begin{acknowledgments}
This work is supported by the Academy of Finland
(Project numbers 104368 and 54023) and by NSERC of Canada (0105354).
\end{acknowledgments}

\end{document}